  \documentclass[12pt,preprint]{aastex}
 \usepackage{ifpdf}
  \usepackage[pdftex]{epsfig}
 \usepackage{amsthm}
 \usepackage{amssymb, latexsym, amsmath}
 \usepackage{amsbsy}
\usepackage{amstext}
\usepackage{hyperref}
\usepackage{amsopn}
\usepackage{amsgen}
\usepackage[dvips]{color}
 \usepackage{graphicx}
  \usepackage{color}
\newcommand{\eb}{\begin{equation}}
\newcommand{\ee}{\end{equation}}
\shorttitle{VIM in Kepler data}
\shortauthors{Makarov \& Goldin}
\begin{document}
\title{Variability induced motion in Kepler data}
\author{Valeri V. Makarov}
\email{vvm@usno.navy.mil} 
\affil{US Naval Observatory, 3450 Massachusetts Ave NW, Washington DC 20392-5420, USA}
\author{Alexey Goldin}
\email{} 
\affil{Teza Technology, 150 N Michigan Ave, Chicago IL 60601, USA}

\date{Accepted . Received ; in original form }

\label{firstpage}
\begin{abstract}
Variability induced motion (VIM) is an observable effect in simultaneous astrometric and photometric measurements
caused by brightness variation in one of the components of a double source or blended image, which manifests
itself as a strongly correlated shift of the optical photocenter. We have processed the entire collection of
the Kepler long-cadence light curve data looking for correlated signals in astrometry and photometry on the
time basis of a quarter year. Limiting the VIM correlation coefficient to 0.3, VIM events are detected for $129\,525$
Kepler stars at least in one quarter. Of 7305 {\it Kepler objects of interest} (KOI), 4440 are detected as VIM at least once. Known variable stars and resolved double stars have elevated rates of VIM detection. Confident VIM
occurrences are found for stars with suggested superflare events, indicating possible signal contamination.
We present a complete catalog of all quarterly VIM detections. This catalog should be checked for such astrophysically
significant events as transits of exoplanets, new eclipsing stars, and superflares of solar-type stars.

\end{abstract}


\section{Introduction}
\label{intr.sec}
Detection of exoplanets based on the photometric data collected by the {\it Kepler} mission is a multi-staged
process, where identification of possible planet transits is just a first step toward a credible and
sufficiently confident science case. The extended and laborious verification
schema includes a search for nearby
stars for each candidate Kepler object of interest (KOI) using high-resolution and adaptive optics ground-based instruments.
The need for this search emerges from the combination of superb photometric sensitivity of the Kepler instrument
with a rather poor angular resolution and optical image quality. 

The telescope is of a Schmidt design with a large field of view of 115 deg$^2$ \citep{koc}, requiring a corrector plate and 
a separate field-flattening assembly of lenses. The detector includes 42 CCDs divided into 84 separate
read-out channels.
Each pixel of 27 $\mu$ on a side covers $3\farcs 98$ on the sky. The optical
point-spread function (PSF) is rather broad with a 95\% encircled flux radius
of 16 to 28 arcseconds. The minimum diameter of the PSF is 4.4 pixels mostly defined by the Schmidt corrector.
Even widely separated double stars and optical pairs have a substantial amount of light
blended in the pixel apertures transmitted to the ground. The effective pixel response function (PRF), which includes
pointing jitter and the optical PSF, may acquire strongly asymmetric shapes toward the edges of the field of view
\citep{bry}. The electronic apertures used for downloading the data, photometry, and astrometry, are also of irregular,
asymmetric shape. The configuration of overlapping images for a fixed pair of sources may therefore depend on the orientation of the
CCD on the sky. A double source can be detected as VIM in one quarter ($\sim 90$ days) but not in the
next because of the different orientation of the pixel grid on the sky. 

\citet{cou} provided a comprehensive analysis of the different types of signal blending and contamination,
which can lead to a detectable VIM effect if at least one of the sources is variable. Double sources are not
the only reason for VIM effects. Ghost images of brighter stars may appear within $\sim 40$ arcsec due to reflections
of starlight off the field flattener lenses, and across the entire field of view due to reflections off the corrector
plate. A distinctly different mechanism of contamination is caused by electronic interferences in the CCD readout,
when a signal charge in one channel may electromagnetically induce a spurious signal at similar pixel coordinates
in the three adjacent output channels. The range of this contamination is a fraction of the field of view. A still
unexplained ``column anomaly" manifests itself as an entire set of pixel columns (along which the charge is read out)
being illuminated by a star. A variable source, which may not be a Kepler object, can be blended with all Kepler
targets with close pixel $x$-coordinates. The range of this mechanism is the extent of a CCD. Bright stars also cause
a large-scale checkerboard pattern called moir{\'e} \citep{cal}.
 
Rigorous procedures are in place to vet detected variations in Kepler light curves that may be caused by transiting exoplanets
\citep[e.g.,][]{bor, bat} and to reject false positives. One of the verification stages includes a test for correlated
photocenter shifts at the pixels level, in and out of the transit \citep{bry13}, which is essentially a VIM detection
in the terminology of this paper. In many cases, follow-up observations are required to confirm low-SN ratio cases.
This level of rigor and caution has been applied to candidate exoplanet events, designated
as KOI. The broad variety of other astrophysical phenomena where relatively small or
infrequent light variations are important, receive less scrutiny, and may suffer from a higher rate of false
positives. We adopted a different, rather empirical,  approach in this paper attempting to detect VIM effects
for {\it the entire set} of Kepler objects, and all the 17 seasons of the primary mission. This results in a massive catalog of
potential light contamination events, irrespective of possible causes.

The motion of star images across the pixel grid caused by the differential aberration of light, residual pointing
drift, and micrometeoroid strikes is a common source of correlated systematic errors in Kepler photometry and astrometry.
As the distribution of light shifts with respect to the fixed digital aperture, different portions of the image
get clipped in the wings, changing both the total aperture flux and the flux-weighted centroid. This results in
a Motion-induced Variability, or MIV, effect, which is the inverse of the VIM investigated in this paper. MIV should be
omnipresent in the unfiltered aperture data, especially closer to the edges of the field of view, 
because it occurs even for intrinsically constant stars. Most of the
MIV occurrences are long-term, however, affecting the data on time scales of a quarter. The technique described in Section
\ref{method.sec} is designed to minimize the rate of MIV detections, but a clear distinction is not possible, especially when long-term intrinsic variability of stellar flux is concerned.

\section{Objectives and disclaimers}
We include an object in our catalog whenever we detect a significant correlation between the light curve and the 
astrometric trajectory in pixel coordinates. This does not make a distinction of {\it parent} and {\it children}
sources in the terminology of \citet{cou}. In other words, it remains unknown which of the blended sources
is variable, or even where the variable source is located with respect to the target. For regular VIM in double
stars, including unresolved binaries, the observable parameters are insufficient to unambiguously determine the
separation between the blended sources \citep{pou, per}. Only the lower limit of separation can be estimated,
and the position angle of the constant component with respect to the variable component. Our results indicate that
a large fraction, possibly the majority, of detected Kepler VIM events are caused by long-range blending of signal,
so this analysis is of limited value for detection of unresolved binaries. Owing to the superb short-term photometric
(20 -- 40 ppm) and astrometric ($<1$ mas for brighter stars) single-measurement precision, new unresolved binaries
should be present in our sample, but the available data is insufficient to identify them.

\section{Example of VIM in Kepler data}
\begin{figure}[htbp]
\plotone{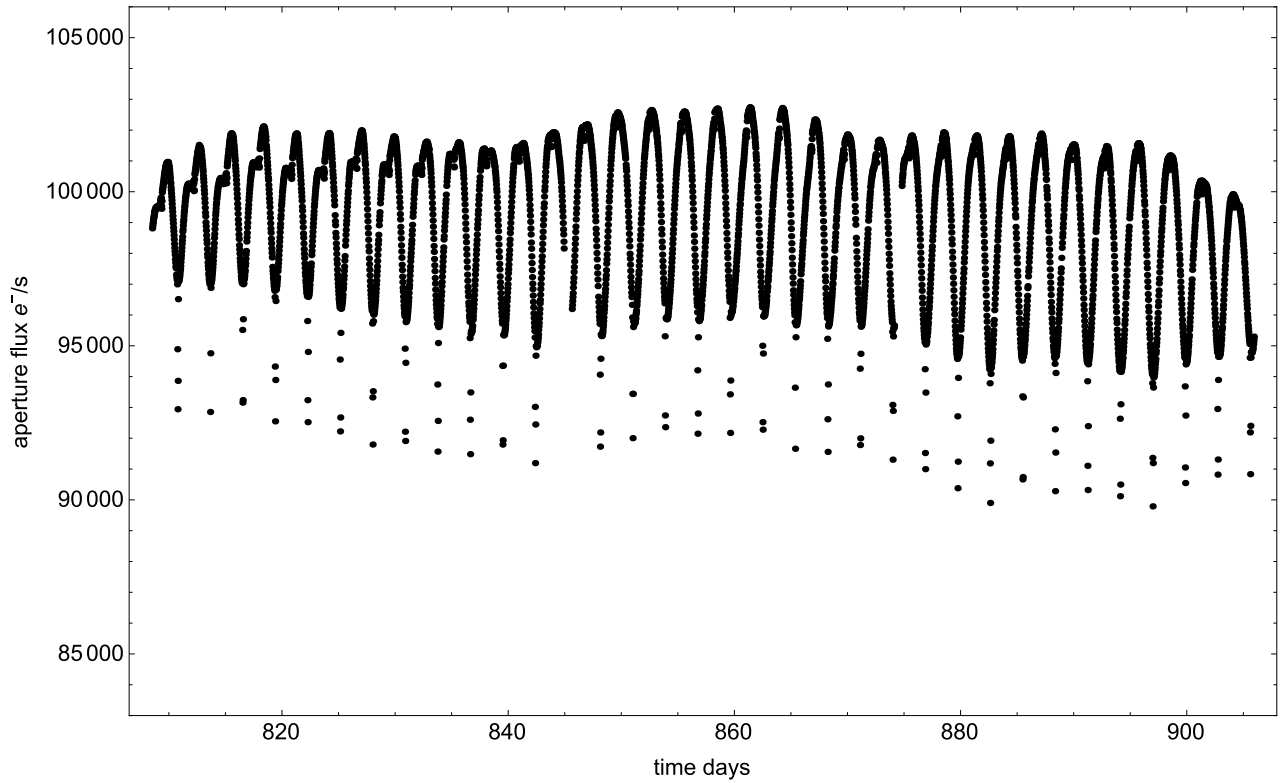}
\hspace{2pc}
\caption{A segment of Kepler aperture light curve for the eclipsing star KIC 4035675.}
\label{light.fig}
\end{figure}

\begin{figure}[htbp]
\plotone{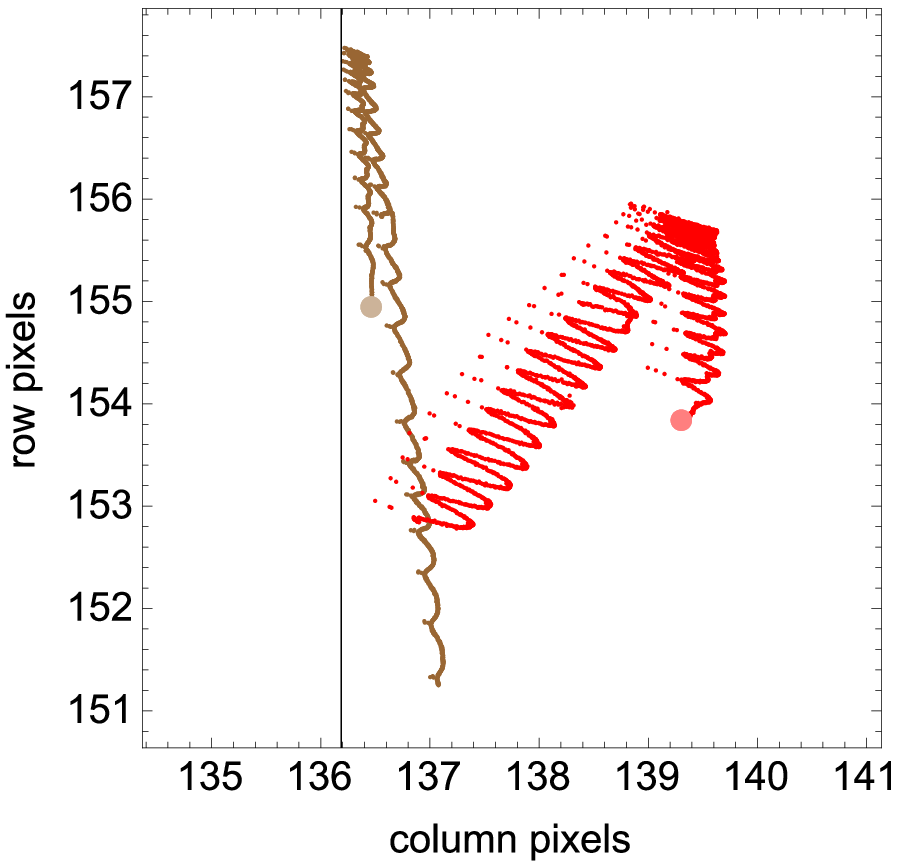}
\hspace{2pc}
\caption{Detected centroid trajectories of two nearby Kepler targets, KIC 4035675 (red) and 4035667 (brown). The offsets
from the starting positions indicated by large circles, are magnified $\times 25$ to the scale of the plot.
}
\label{2stars.fig}
\end{figure}

The star KIC 4035675 at RA$=284.62463$, Dec$=+39.17878$ degrees is an eclipsing binary of Algol type with a
period of 2.87 days \citep{sla,prs}\footnote{Also see the elegantly arranged and useful online Villanova catalog
of eclipsing binaries at {\tt keplerebs.villanova.edu}}. It happens to be located close to another, much brighter
star TYC 3119-347-1, which is also a Kepler target KIC 4035667, possibly a variable of $\delta${\it Sct} type
\citep{uyt}. The angular separation between these two stars is $\sim 13.7$ arcsec. The digital apertures for the
stars, as seen from the archival target pixel data, are of irregular shape and large enough to allow for a
considerable amount of blending to occur. The much stronger, strictly periodic, variation of the eclipsing component
shown in Fig. \ref{light.fig} dominates the light curve of its neighbor. The shifting secondary dip indicates variability
with a slightly longer period, probably caused by a long-term spot on the surface of the star rotating at a subsynchronous
rate.
Both astrometric trajectories also show a periodic pattern very similar
to the light curve. Fig. \ref{2stars.fig} shows the centroid positions of both stars in the first half of
Quarter 9. The initial positions in pixel coordinates are indicated with large filled circles. The deviations of
centroids recorded in the long-cadence files are shown with dots, magnified by a factor of 25 to the scale of the plot,
to visualize the periodic wiggles.

The centroid trajectory reproduces extremely faithfully the light curve of the eclipsing component. The detected
centroid of light is shifted from the true image location by the contribution of light from the brighter neighbor.
This contribution can be considered practically constant for this analysis, whereas the bulk of the detected
light coming from the intended target varies in a wide range because of the eclipses. The photocenter of the blended
light shifts synchronously along the line connecting the two sources. A symmetric, but much weaker,
effect is observed in
the aperture of KIC 4035667 (Fig. \ref{2stars.fig}). Note that the direction of astrometric deviations 
in either trajectory corresponds to the position angle
of the double star configuration in this case. The detail of the VIM trajectory is remarkable owing to the superb
astrometric short-term, single-measurement precision (of order 1 mas for this source), reproducing quite clearly
the shifting ``secondary eclipse". 

\section{Method}
\label{method.sec}
The primary data used in this analysis are the moment-based centroid positions in pixel coordinates (\verb|MOM_CENTR1|
and \verb|MOM_CENTR2|) given for each individual cadence step at time (field {\verb|TIME|) and 
the corresponding aperture fluxes (\verb|SAP_FLUX|) in e$^{-}$ s$^{-1}$ from the long-cadence light curve (llc) files available through
the MAST Kepler archive\footnote{A description of the archive is available at \newline
\url{http://archive.stsci.edu/kepler/manuals/archive_manual.pdf}}. The observing strategy of Kepler was to 
keep the same pointing and orientation
of the telescope as accurately as possible for about 90 calendar days of continuous observations. Still,
the centroids of star images could move by hundredths to sometimes tenths of a pixel during each quarter
because of the differential aberration of starlight and some other instrumental effects of yet unknown origin
\citep{ben}. 
A typical centroid trajectory as a function of time for star KIC 3974043 is shown in  Figure \ref{drift.fig}.
Our first task is to remove the ``creep", i.e., the slowly varying part of the trajectory. This has to be done
because the likely reason for the creep is not a long-term stellar variability of blended signals (such phenomena
are relatively rare) but the detrimental technical effects, which may not be related to VIM.
\begin{figure*}
\includegraphics[width=150mm]{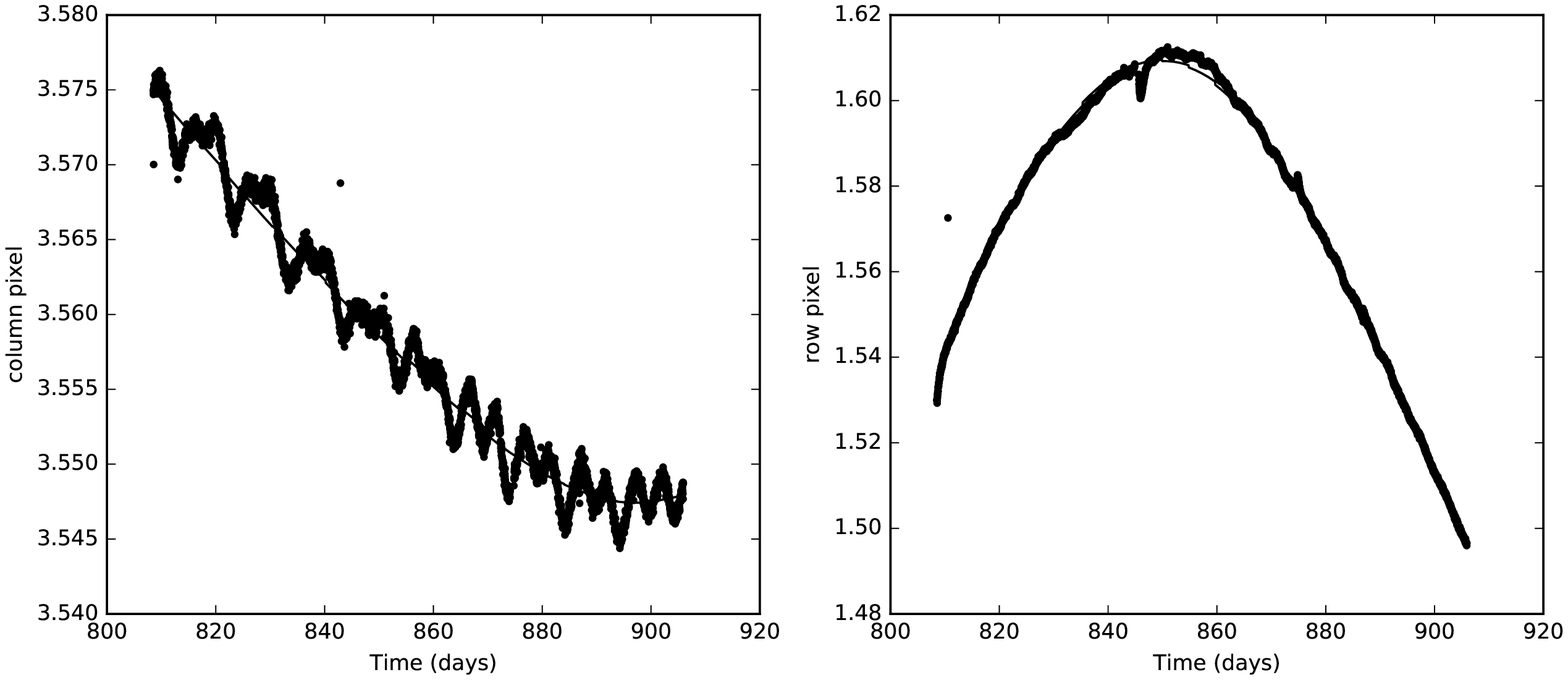}
\hspace{2pc}
\caption{Example of measured centroid position drift in chip coordinates for star KIC 3974043.}
\label{drift.fig}
\end{figure*}

The relatively large motion observed in the raw moment-based centroids such as the one shown in Figure \ref{drift.fig}
is mostly caused by the residual pointing drift and roll of the telescope, and the effects of differential aberration
of light within the large field of view. In particular,
the differential aberration arising from the barycentric velocity of the space craft of approximately
30 km s$^{-1}$ accounts for a significant fraction of the systemic long-term motion of star images
on the came detector channel, reaching tenths of arcsecond for stars farther from the field center. 
However, we determined by extensive experiments with the data that when the aberration component is removed 
by direct modeling of pixel trajectories
or by affine transformations for each channel, a large residual trajectory remains (astrometric creep), which is not
correlated with the position on the chip or pixel phase. This residual creep
looks stochastic for each star and has a typical magnitude  of $\sim 0.1$ arcsecond, which is 2--3 orders of magnitude
greater than the single measurements astrometric precision of the long-cadence centroids. 
The pipeline-provided centroids
\verb|POS_CORR1|, \verb|POS_CORR2| do not improve the situation from our point of view, because they still suffer from the large creep
of unknown origin. We chose to use the original flux-weighted centroids and to de-trend them by a polynomial
fitting. Fitting a single linear or parabolic model to the trajectory did not work well enough. 
Our
solution to filtering out the low-frequency component from the positions is to use the local regression (LOESS),
where at each data point we fit a local low degree polynomial model \citep{1979,1988}.
The LOESS filtering removes the long-term trends in positions quite efficiently, including the
aberration tracks and the residual creep.

Specifically, to fit the coordinate at time $t$ the weight of the model is 
\begin{equation}
  \label{eq:wloess}
  w(t)  =  \left( 1 -  \frac{\left| t - \tau \right| } {\Delta t}\right) ^3
\end{equation}
where $\tau$ is the running time variable, $|t-\tau|<\Delta t$.
A new two-degree polynomial is fitted for each point in time. $\Delta t$ is the only parameter that has to be adjusted. 
We typically set it to $\approx 0.1$ of the single track (90 days long). The interval of LOESS defines the effective
bandpass of the filter. We found by numerical simulation that with an interval of 0.1 and the standard second degree
polynomial, and for a sinusoidal signal, approximately 20\% of power is lost in LOESS filtering at 0.1 period,
and 50\% at 0.14 period. This smoothing property makes our method insensitive to slower 
variability induced motion, which is the cost of suppressing the omnipresent long-term instrumental effects. The effective
time scale of suppressed variability is about 9 days and longer.

After the fitted LOESS curve is  subtracted from the observed trajectory, the typical residuals 
(with the slow time-dependent component removed) are shown in figure \ref{fig:Resid}. The trajectory in the 1st coordinate,
i.e., the column pixel on the CCD, displays a systematic periodic motion with an amplitude of $\sim 12$ mas and
with a period of roughly 10 days. This systematic motion is caused by the true rotational variability of the star
\citep{deb,mcq}, and is not a measurement error. 
The residual row pixel trajectory still includes some jumps and dips (right panel), probably of instrumental origin,
but the dramatic creep across $\sim 700$ mas is eliminated. The 10-day periodic is still visible but with a much smaller
amplitude. The next step is to calculate to what degree the systematic pattern in residual coordinates is correlated to
the light curve.
\begin{figure*}
\includegraphics[width=150mm]{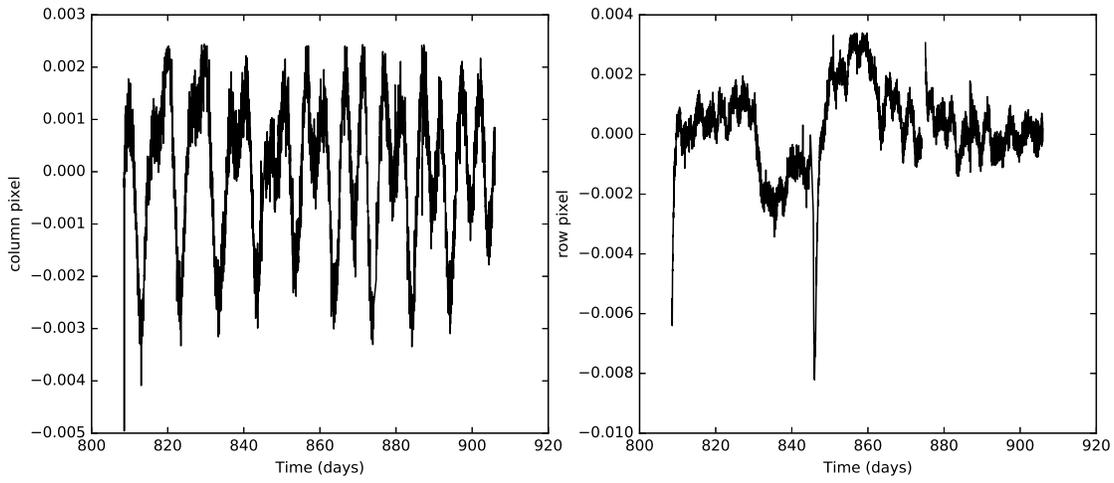}
\hspace{2pc}
  \caption{Residual column (left) and row (right) coordinates after the slow part of centroid drift is subtracted}
  \label{fig:Resid}
\end{figure*}

Residual coordinates versus flux are shown in figure \ref{fig:Resid_Flux}. While the column coordinate is obviously correlated with the flux, this correlation is much less noticeable in the row coordinate. A few very obvious outliers are also present.
\begin{figure*}
\includegraphics[width=150mm]{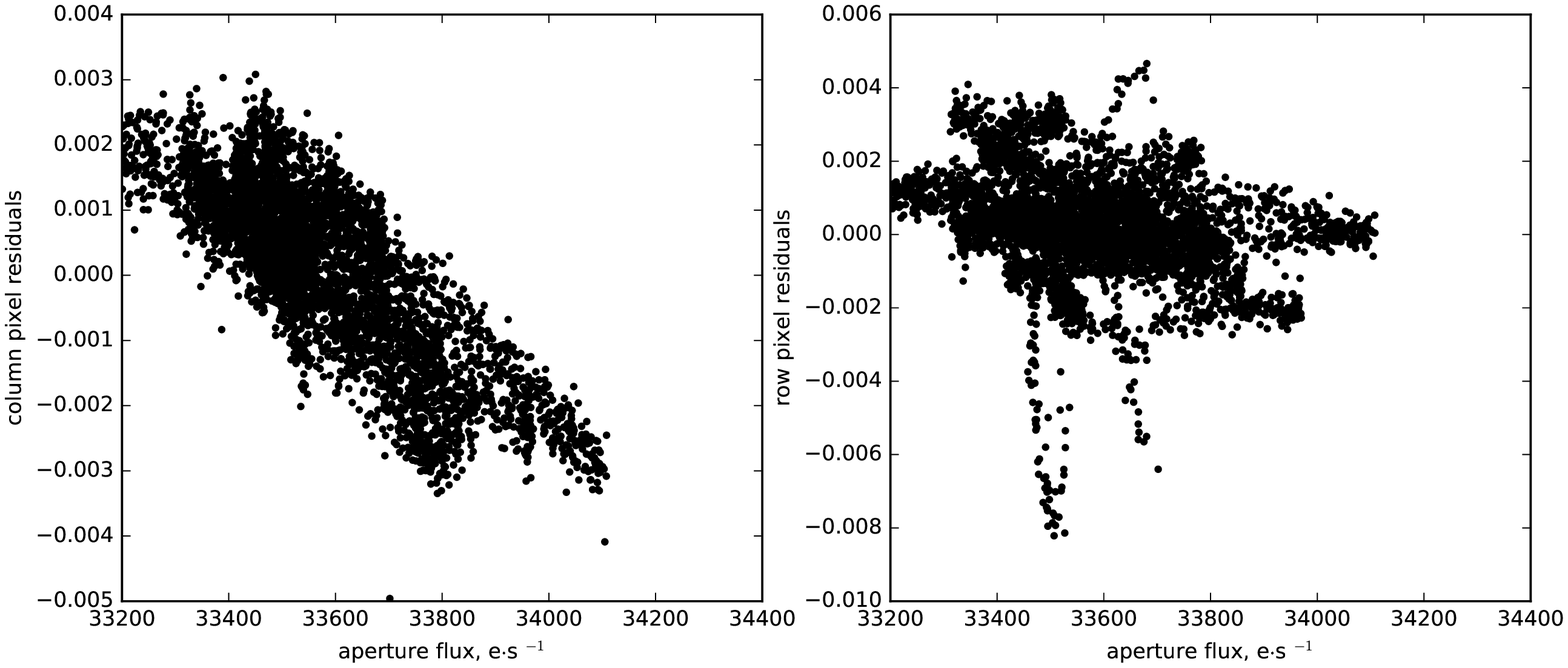}
\hspace{2pc}
  \caption{Residual coordinates as functions of flux}
  \label{fig:Resid_Flux}
\end{figure*}
We fit a simple linear model 
\begin{eqnarray}
  \label{eq:1}
  s_1 = a_1 f + c_1 \\
  s_2 = a_2 f + c_2
\end{eqnarray}
where $f$ is the flux, $s_1$ and $s_2$ are the chip coordinates (residuals). To guard against outliers evident in figure \ref{fig:Resid_Flux}, we use the robust linear regression with the Tukey bisquare weight function \citep{Mosteller:1977:DAR, statsmodels} instead of the ordinary least squares. As a by-product, the robust linear regression generates a vector 
of weights $w_i$, which get close to zero for detected outliers. In the calculation of position-flux correlation,
we simply neglect all data points with the lowest 5\% weights. 

At this point we can find a rotation in the coordinate plane which maximizes the position-flux correlation in the first (column) coordinate. In other words, the unit vectors of the new coordinate system are chosen in
such a way that the variance induced motion becomes the largest in the column coordinates:
\begin{eqnarray}
  \label{eq:rotation}
  \vec{u_1} = \left( \frac{a_1}{\sqrt{a^2_1 + a^2_2}}, \frac{a_2}{\sqrt{a^2_1 + a^2_2}}, \right) \\
  \vec{u_2} = \left( \frac{-a_2}{\sqrt{a^2_1 + a^2_2}}, \frac{a_1}{\sqrt{a^2_1 + a^2_2}}, \right).
\end{eqnarray}
 This rotation specifies the position angle of the VIM, i.e. the direction in which the detected coordinated motion
 takes place relative to the average target position. This value may indicate the locus of a close blended source,
 for example, a double star companion.

 \begin{table*}
 \centering
 \caption{KepVIM catalog description.}
 \label{format.tab}
 \begin{tabular}{@{}lrr@{}}
 \hline
            &                 &      \\
   Name     &  Description    & Units \\
 \hline
 $KEPID$ & \dotfill KIC number &   \\
 $RA$   & \dotfill right ascention             & deg \\
 $DEC$ & \dotfill declination & deg \\
 $FLUX50$ & \dotfill median flux & e$^-$ s$^{-1}$ \\
 $POSXrange$ & \dotfill 90\% $X$ position range & pix\\
 $POSYrange$ & \dotfill 90\% $Y$ position range & pix\\
 $r$ & \dotfill VIM correlation coefficient & \\
 $AX$ & \dotfill VIM direction cosine $X$ on CCD & \\
 $AY$ & \dotfill VIM direction cosine $Y$ on CCD & \\
 $ds/dF$ & \dotfill VIM speed &  pix\\
 $posA$ & \dotfill VIM position angle on the sky &  degr\\
 $CHANNEL$ & \dotfill detector CCD channel &\\
 $Q$ & \dotfill mission quarter & \\
 \hline
 \label{table}
 \end{tabular}
 \end{table*}

 \begin{table*}
 \centering
 \caption{KepVIM catalog contents (excerpt).}
 \label{cat.tab}
 \begin{tabular}{@{}lrrrrrrrrrrrr@{}}
 \hline
            &                 &      \\
 (1) & (2) & (3) &(4) & (5) & (6) & (7) & (8) & (9) & (10) & (11) & (12) & (13)\\
 \hline

2848980 & 290.69468 & 38.08814 & 8974 & 0.019 & 0.037 & 0.39 & 0.292 & -0.956 & 7.9E-5 & 344.2 & 1 & 2\\ 
2848980 & 290.69468 & 38.08814 & 8698 & 0.003 & 0.003 & 0.50 & 0.541 & -0.841 & 1.1E-5 & 359.8 & 1 & 6\\
2848987 & 290.69589 & 38.05188 & 44402 & 0.001 & 0.001 & 0.35 & -0.228 & 0.974 & 6.7E-7 & 160.3 & 81 & 8\\
2848987 & 290.69589 & 38.05188 & 44051 & 0.002 & 0.002 & 0.50 & 0.657 & -0.754 & 7.2E-6 & 8.2 & 29 & 17\\
2848999 & 290.69772 & 38.01647 & 44798 & 0.003 & 0.003 & 0.31 & 0.597 & -0.802 & 1.0E-6 & 3.7 & 1 & 10\\
2848999 & 290.69772 & 38.01647 & 43193 & 0.002 & 0.004 & 0.36 & 0.723 & -0.691 & 4.7E-7 & 13.3 & 53 & 11 \\
\hline
 \label{table}
 \end{tabular}
 \tablecomments{Columns: (1) $KEPID$, (2) $RA$, (3) $DEC$, (4) $FLUX50$, (5) $POSXrange$, (6) $POSYrange$, (7) $r$, 
 (8) $AX$, (9) $AY$, (10) $ds/dF$, (11) $posA$, (12) $CHANNEL$, (13) $Q$, see Table \ref{format.tab}.}
 \end{table*}

The correlation coefficient $r$ is computed as the absolute value of the sample Pearson's coefficient:
\eb
r=\frac{\left| \sum s_i f_i -n \langle s \rangle \langle f\rangle\right|}{\sqrt{\sum s_i^2-n\langle s\rangle^2}\sqrt{\sum f_i^2-n\langle f\rangle^2}}
\ee
where $s_i$ and $f_i$ are the sampled position (LOESS-filtered and rotated) and flux, respectively, $\langle s \rangle$ 
and $\langle f \rangle$ are their sample mean values, and $n$ is the sample size. The absolute value removes the sign of
the correlation, so that thus computed $r$ values run from 0 to 1. A value of 1 implies a very tight, quite linear
dependence of $s$ on $f$. It is practically impossible to reach it with real data because of the presence of uncorrelated
noise in the measurements, but the calculated $r$ can come close to 1 for most pronounced VIM effects due to the
high relative precision of the photometric and astrometric measurements. The other factor which may significantly
reduce the calculated $r$ values is a long-term variation of instrumental origin in the light curves,
which we chose not to filter. The motivation for this one-sided high bandpass filtering is to minimize the propagation
of the adverse MIV effects, but it may result in somewhat underestimated correlation coefficients for true VIM.

\section{Description of the catalog}

The algorithm described in the previous section was applied to the entire collection of ``long-cadence" files
archived in the MAST for the principal Kepler mission. A single VIM detection corresponds to a complete data set for a given target collected during one quarter. Therefore, a single target can generate up to 17 VIM detections in the
catalog. We selected a uniform threshold on the correlation coefficient $r$ for the published catalog, equal to 0.3.
This was the only criterion for inclusion of a detection in the catalog. The total number of VIM detections in
the catalog is $343\,298$. The number of stars that were detected as VIM in at least one quarter is $129\,525$,
which represents a large majority of the entire sample. The distribution of the number of detections per quarter is
strongly nonuniform, ranging from 5995 for Quarter 5 to $40\,083$ for Quarter 14, possibly indicating the significance
of instrumental causes. The distribution of the number of detections per star is exponentially falling off from
$52\,591$ at 1 (i.e., just one detection out of 17 trials) to 687 at 10, followed by a fairly uniform tail stretching
all the way to 17, where the number of detections is 423. This behavior suggests that the majority of VIM events are
circumstantial, rather than persistent effects caused by a close companion star always present in the target's aperture.

The contents of the KepVIM catalog is given in Table \ref{format.tab}. Each line in the catalog specifies a
single VIM detection for a particular target, identified by the KIC number, in a particular quarter ($Q$).
A few lines extracted from the catalog are shown in Table \ref{cat.tab}. 
Some information about the star ($RA$ and $DEC$) and about the observation ($CHANNEL$) is copied from the
Kepler archive for user's convenience. The other data fields contain parameters derived by our data processing.
The field $FLUX50$ specifies the computed median flux for the quarter. 
The magnitude of astrometric variation after the dynamic fit in the $X$ (column) and $Y$ (row)
coordinates are given as the 90th percentile interval in $POSXrange$ and $POSYrange$. One of the most important
parameters is $r$, the coefficient of VIM correlation. The direction of the correlated
motion in the pixel coordinates $(X,Y)$ is specified by the pair of direction cosines $AX$ and $AY$. The VIM speed
$ds/dF$ determines the magnitude of VIM in pixels caused by a unit change of the total flux $F$. It is determined
from the slope of the linear correlation dependence as shown in Fig. \ref{fig:Resid_Flux}.
The position angle of the VIM in the sky coordinates,
reckoned from the North through East, is given in the $posA$ field. 

We included all tentative detections with $r>0.3$ trying to minimize the presence of the inverse instrumental
effect, MIV, as described in Section \ref{intr.sec}. This effect results in long-term correlated variations in
detected positions and flux even in the absence of intrinsic stellar variability. It is common, but of little interest
to researches investigating short-term phenomena such as exoplanet transits or short-period eclipsing binaries.
Setting a high threshold for $r$ is meant to preferably select objects with a higher degree of intrinsic variability.
To estimate the statistical significance of detection, we use a bootstrap method on the real long-cadence data for
all Kepler targets in one quarter (Q5). For each star, the centroid trajectories and the SAP light curve are
processed as described in Section \ref{method.sec} and the $r$ value is determined. The trajectories and the
light curves are then randomly scrambled (reordered in time) 1000 times and the correlation coefficient is
calculated each time using the same algorithm. The p-value is estimated as the number of trials with $r$ greater 
than the observed coefficient
from the actual data divided by 1000. The p-value defines the probability of the null hypothesis to be rejected
while being true. A low p-value means that the chance of getting the observed correlation accidentally is
small. Conversely, unity minus p-value defines the confidence of correlation detection given the data. 
Fig. \ref{pvalue.fig} shows the results at a resolution of 0.001 in p-value. Detections at $r>0.3$ included
in the catalog have a confidence level greater than 0.999. A 0.99 level of confidence is achieved already at
$r\simeq 0.08$.

\begin{figure}[htbp]
\plotone{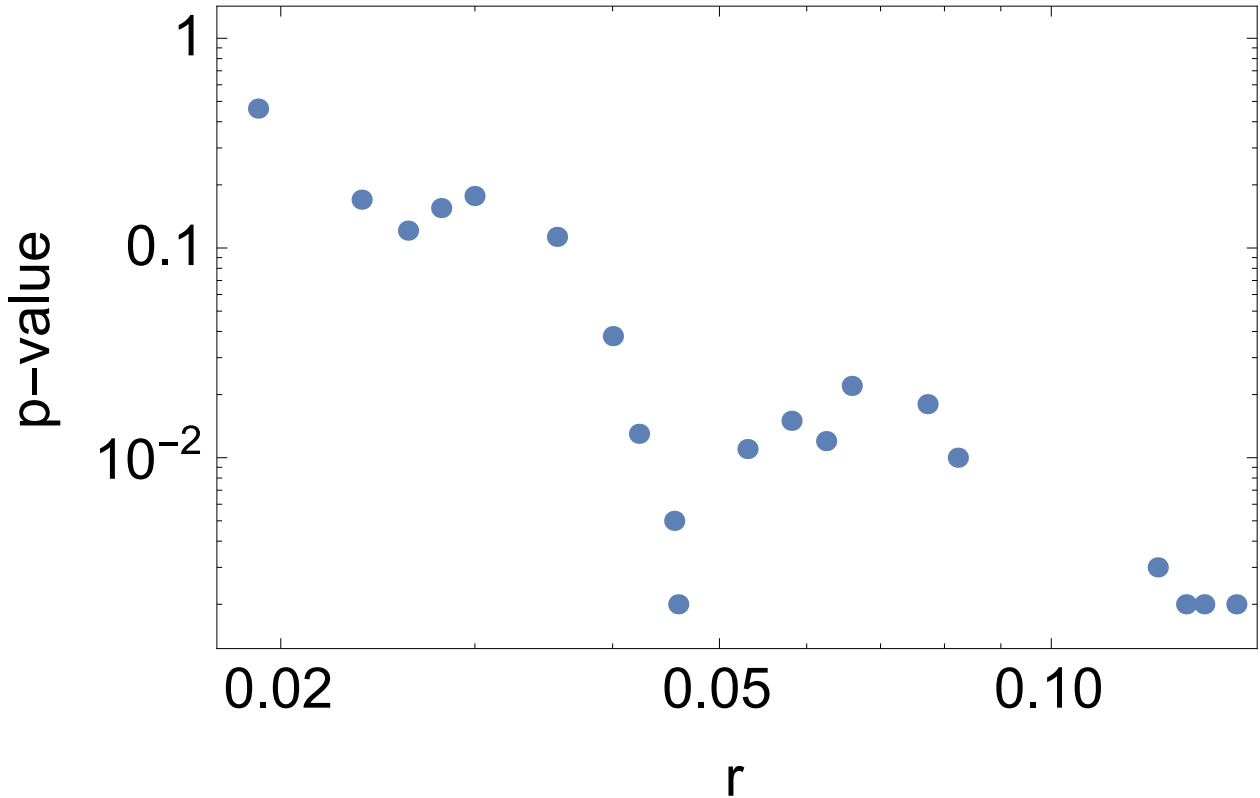}
\hspace{2pc}
\caption{Correlation coefficient p-values estimated by the Monte-Carlo bootstrap method for Kepler data in Q5. 
}
\label{pvalue.fig}
\end{figure}

Some light on the source of detected VIM effects can be shed by the calculated values $AX$ and $AY$. A VIM caused
by an unresolved double can have arbitrary direction on the CCD, because any position angle of a double star
is equally probable. The observed distribution of direction cosines, however, is nonuniform. Fig. \ref{hist.fig}
shows the histogram of VIM position angles for all detections in Quarter 13 in corresponding CCD coordinates.
A clear excess in the distribution around $posA=0\degr$ and $180\degr$ is seen. These angles correspond to the
left and right directions along the rows of pixels on the CCD. There is little doubt that the excess of VIM
detections is caused by the ``column anomaly" long-range effect, cf. \S \ref{intr.sec} and \citep{cou}.
From the relative number of detections in the excess peaks of the histogram, we estimate that up to 28\% of all 
our detections
can be caused by this detrimental effect.

The ability of Kepler to detect unresolved double stars is an enticing possibility. Assuming that the two blended
star images are well within the astrometric and photometric apertures, the photocenter is located at the center
of light of the combined image:
\eb
\frac{f_1}{f_2}=\frac{s_2}{s_1},
\ee
where $f_1$ and $f_2$ are the fluxes of the two components, and $s_1$ and $s_2$ are their separations from the
photocenter. If $s=s_1+s_2$ is the total separation between the components (which is invariable on the sky within
the span of observational data), and only one of the components, say number 1, is variable, the separation can be estimated
by the simple formula
\eb
s=ds\frac{f}{f_2}\frac{1+\Delta}{\Delta},
\label{s.eq}
\ee
where $ds$ is the magnitude of astrometric VIM excursion, $f=f_1+f_2$ is the total median flux, and $\Delta=df/f$ is the magnitude
of photometric variation correlated with the astrometric excursion. The ratio $ds/\Delta$ can be computed from the
catalog as 
\eb
\frac{ds}{\Delta}=ds/dF \times FLUX50,
\label{sep.eq}
\ee
i.e., it is the VIM speed times the median total flux. 
\begin{figure}[htbp]
\plotone{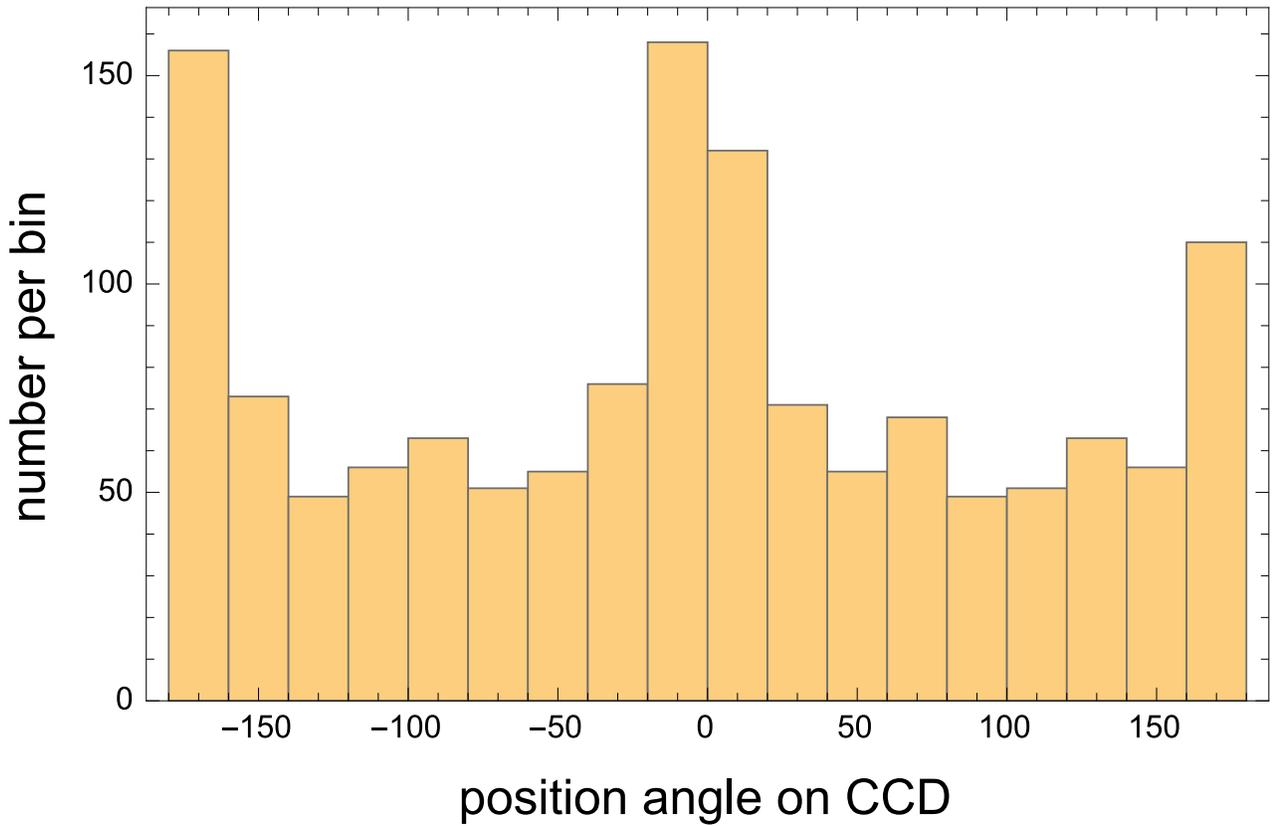}
\hspace{2pc}
\caption{Distribution of detected VIM position angles in pixel coordinates for Quarter 13.}
\label{hist.fig}
\end{figure}

Equation \ref{s.eq} suggests that double systems with
components of equal brightness can be detected down to separations of $\sim 80$ mas if one of the components is
variable at 10\% of its median flux, because for many stars, Kepler astrometry is precise enough to detect excursions
of $\sim 0.001$ pixel, or $4$ mas on the sky. Most of the test cases we have considered do not live up to these
expectations. For example, the star KIC 5557932 = HD 226195 = HIP 97706 is an interesting solar analog (G1.5V)
with a Hipparcos-determined parallax of $13.52\pm 0.79$ mas. Hipparcos also found a ``C"-type solution for this
object, resolving it into two components separated by $0\farcs 315$ on the sky at a position angle $\theta=346\degr$.
The Kepler light curve betrays a complex variability pattern, possibly due to a fast rotation rate \citep{nie}. 
Our catalog lists 15 detections out of 17 with correlation coefficients $r$ reaching 0.94 in Q6, indicating a
confident and persistent VIM effect. The VIM position angle was detected as $posA=334\degr$ in Q3, $336\degr$ in Q7,
$353\degr$ in Q11, and $327\degr$ in Q15, with a mean  of $337.5\degr\pm 11.0\degr$. The other detections were
$posA=155\degr$ in Q1, $153\degr$ in Q4, $151\degr$ in Q5, $150\degr$ in Q6, $150\degr$ in Q8, $153\degr$ in Q9,
$150\degr$ in Q10, $149\degr$ in Q12,  $151\degr$ in Q14, $152\degr$ in Q16, and $151\degr$ in Q17, with a mean  of $151.4\degr\pm 1.7\degr$. Note that the difference between the two batches of $posA$ determinations is $180\degr$
within the statistical error. Such switching between two opposite directions may be caused by blended sources of
approximately equal brightness projected on pixels of unequal sensitivity, or essential variability of both
companions. Furthermore, the Tycho-2 photometric data reveal components of a $0.5$ magnitude difference,
but very close in $B_T-V_T$ color \citep{fab}, which may be indicative of a hierarchical multiple system.
The first batch of detections corresponds to a specific configuration of the sources on the pixel grid,
because they are separated by four quarters. There is little doubt that both batches come from the same
double source with a position angle of $331\degr$. This is close to the Hipparcos value\footnote{The small difference in
the detected position angle may well be due to the orbital motion in the intervening 20 years between the two
missions}, so the case may appear to be a true positive detection of a tight double star. The crucial difficulty
comes from the interpretation of the VIM speed, $ds/dF$, data. Using Eqs. \ref{s.eq} and \ref{sep.eq}, we can estimate
the {\it minimum} separation between the components, which is also switching between $\sim 0.23$ pix for the first
batch of detections, and $\sim 0.90$ pix for the second. This indicates that the measured VIM speed strongly depends
on the pixel configuration. In either case, the estimated separation is too high compared to the Hipparcos
measurement, by a factor of 10 or greater. We do not have an explanation to this perplexing magnification of VIM
in the Kepler data.

\section{Objects of note}
\subsection{KOI}
At the time of writing this paper, the Kepler archive included 7305 {\it Kepler objects of interest} (KOI). These
stars are the ones with possible exoplanet transits in the light curves determined by the preliminary analysis. 
The intrinsic ambiguity whether the detected dip in the light curve actually comes from the intended target is hard
to resolve within the collected mission data. A VIM detection in the same quarter as a tentative transit
detection indicates a confidently detected blending occurrence. The blended source may be an eclipsing star,
whose light is diluted in the aperture of the intended target reducing the depth of the eclipse and thus
mimicking an exoplanet transit. Using mean densities of host stars determined by asteroseismology,
\citet{sli} suggested that 70\% of confirmed giant exoplanets could be
false positives caused by signal blending. \citet{san} estimated a false positive rate of $54.6\pm6.5\%$
for transiting giant planet candidates by radial velocity follow-up observations. These recent finds perhaps
call for a revision of signal blending contamination in the transit detection technique. We find 4440
KOI with at least one VIM detection in the catalog. However, only 321 KOI are flagged as ``confirmed" in field 5
of the KOI list, with another 1946 flagged as ``candidate" in field 5 generated from NExScI analysis. All the 321
objects with VIM detections and ``confirmed" in field 5 are flagged ``candidate" in field 4. This difference
is caused by different verification strategies adopted by the two teams processing Kepler exoplanet data. We recommend
taking additional precautions with VIM-detected KOI before assigning them the ``confirmed" status. Such events,
if they occur in the same quarters as the suggested transits, can be further analyzed using the VIM speed parameter
$ds/dF$, the correlation $r$, the direction cosines $AX$ and $AY$, the sky position angle $posA$, correlating
them with similar VIM detections in the same channel.

Candidate KOIs detected as VIM in our catalog can certainly be legitimate exoplanet transit sources, because signal
blending affects all kinds of objects indiscriminately. The main point of exoplanet verification in this case is
to make sure that the transits are not coming from the long-range instrumental contamination or nearby companions
whose signal may leak into the target's aperture. Fig. \ref{757.fig} shows a small segment of the light curve
collected for the star KIC 757450 = KOI 889 = Kepler 75 in Q9. There are two prominent occultation events at days 827.5 and 836.4,
caused by the periodic passage of a giant exoplanet. Outside the transits, the light curve displays a complex variability
pattern. \citet{heb} confirmed the existence of the planet combining the Kepler observations with a spectroscopic
follow-up using the SOPHIE and HARPS-N spectrographs. They determined a transit period of 8.88 d, and an additional
periodic variation of $\sim 19.2$ d, possibly caused by huge spots on the rotating surface. The combination of
radial velocity and brightness data allows the authors to estimate the mass of the planet at $9.9\pm 0.5$ $M_{\rm Jup}$,
the radius at $1.03\pm 0.06$ $R_{\rm Jup}$, and an orbital eccentricity of $e=0.569\pm 0.01$. There are two lines in our
KepVIM catalog for this star with detections in Q11, $r=0.35$, $posA=27\degr$, and Q15, $r=0.31$, $posA=30\degr$. First we
note that the star is essentially variable, with deep transits occurring every 8.88 days, but VIM is detected in only
2 out of 17 quarters with $r>0.3$. Fig. \ref{757pix.fig} shows the corresponding de-trended
trajectories for the same time span as in Fig. \ref{757.fig}. A slight 3-day periodic wobble may be present in the
trajectory, possibly coming from a weak blended companion, but there is no resemblance to the light curve, nor any
evidence of the transits. Both VIM detections have approximately the same direction. The DSS sky map of the surroundings
reveals a few faint optical companions, with the closest and relatively brighter companion at $\sim 9"$,
position angle $210\degr$\footnote{We remind the reader that the recorded $posA$ may differ by $180\degr$ from the
actual position angle of a double star, depending on which of the blended components is variable.}. This companion is
the likely culprit in our VIM detections, but it has nothing to do with the interpretation of the exoplanet transits. 

\begin{figure}[htbp]
\plotone{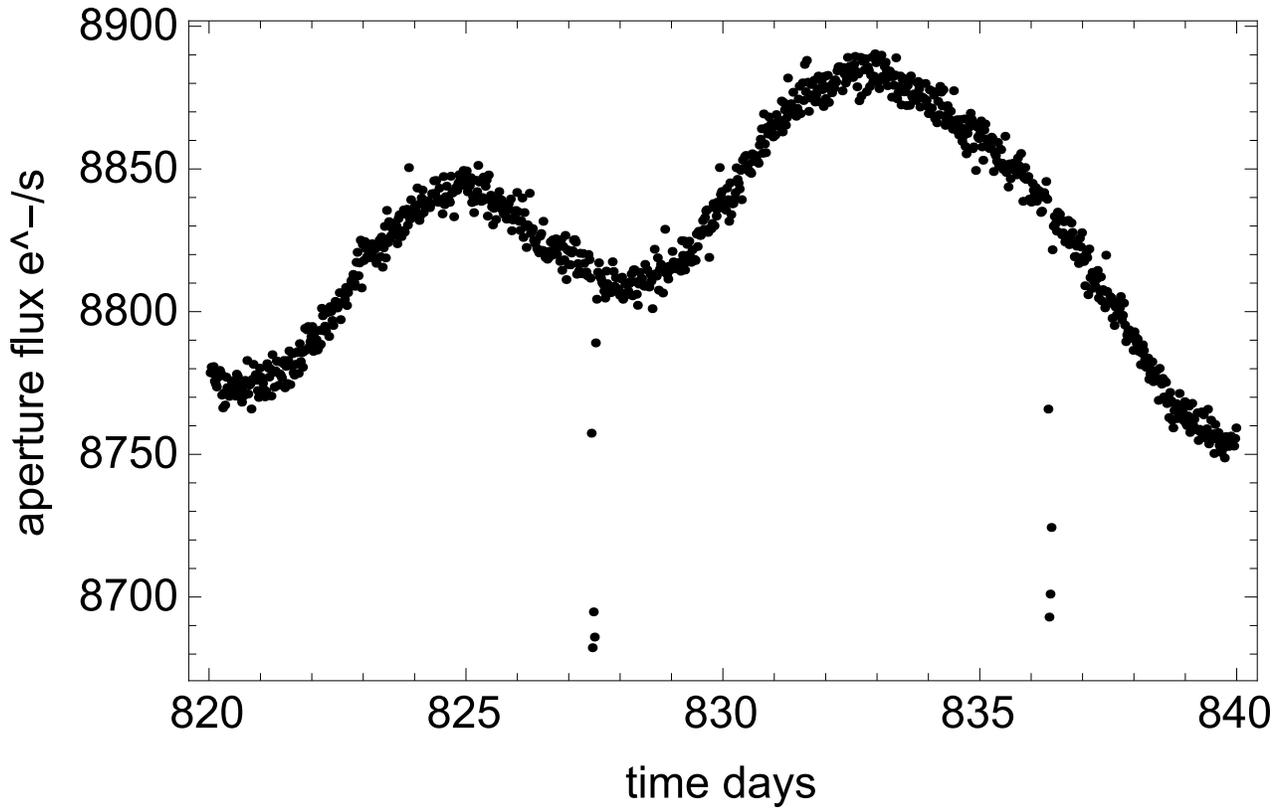}
\hspace{2pc}
\caption{Kepler light curve for the star KIC 757450 with a super-Jupiter giant planet for a span of 20 days between
mission days 820 and 840. Two of the exoplanet transits are clearly seen. 
}
\label{757.fig}
\end{figure}
\begin{figure}[htbp]
\plottwo{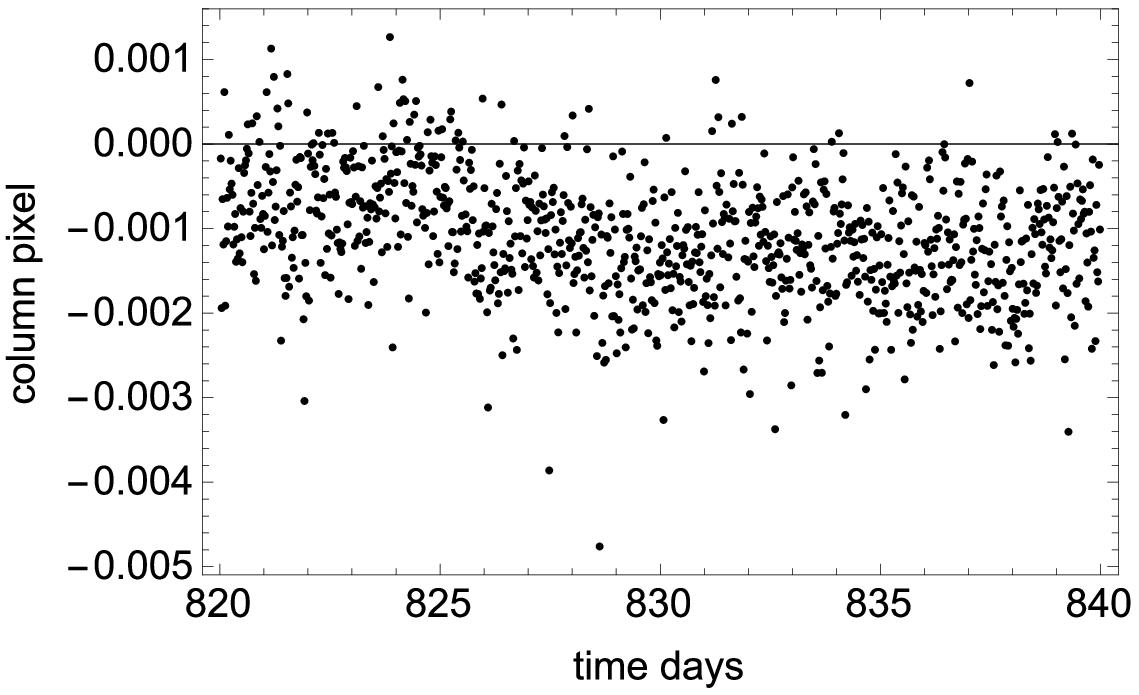}{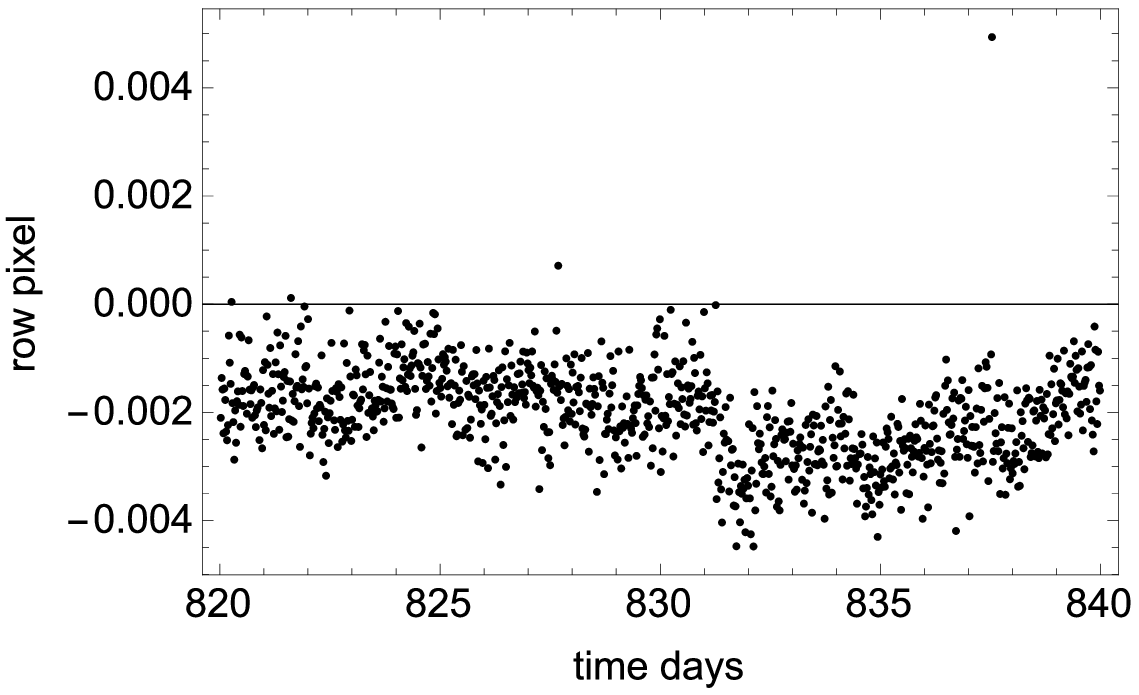}
\hspace{2pc}
\caption{Segments of the long-cadence centroid trajectory in column (left) and row (right) pixel coordinates 
for the star KIC 757450
from Quarter 9 data. No clear
correlation is seen between the brightness variation in Fig. \ref{757.fig} and the centroid position.}
\label{757pix.fig}
\end{figure}

\subsection{Solar-type stars with superflares}
Following the work by \citet{mae}, \citet{wic} investigated a sample of 11 Kepler targets, which seem to be fairly
close solar analogs in terms of their surface gravity and effective temperature, yet displaying powerful spikes
in the recorded light curves, interpreted as ``superflares" analogous to the solar flares that are well-studied 
and extremely important
for humankind. Unlike even the most powerful solar flares, these spikes detected by Kepler can
be as high as 0.1 -- 1.0\% of the total flux. A flare of that magnitude, if it happens on the Sun, could be catastrophic
for the biological life on Earth. The study by \citet{wic} has not considered the possibility of signal contamination,
and our results provide an easy fix to this shortcoming.

All 11 superflare candidates are present in the KepVIM catalog. Three stars, KIC 7264976, 11073910, and 11972298
have been detected as VIM only in one quarter. The largest number of detections, 10 out of 17, is found for KIC 
9653110. The correlations $r$ range from the moderate 0.3 -- 0.4 to the highest values of 0.92 for the star 11610797
in Quarters 7 and 15. The rate
of high $r$ values suggests that these objects are more variable than the average and/or are subject to a greater
degree of signal blending. Within the scope of this study, we will limit our analysis to the first star in
the sample, KIC 3626094 = TYC 3119-1230-1. 

\begin{figure}[htbp]
\plottwo{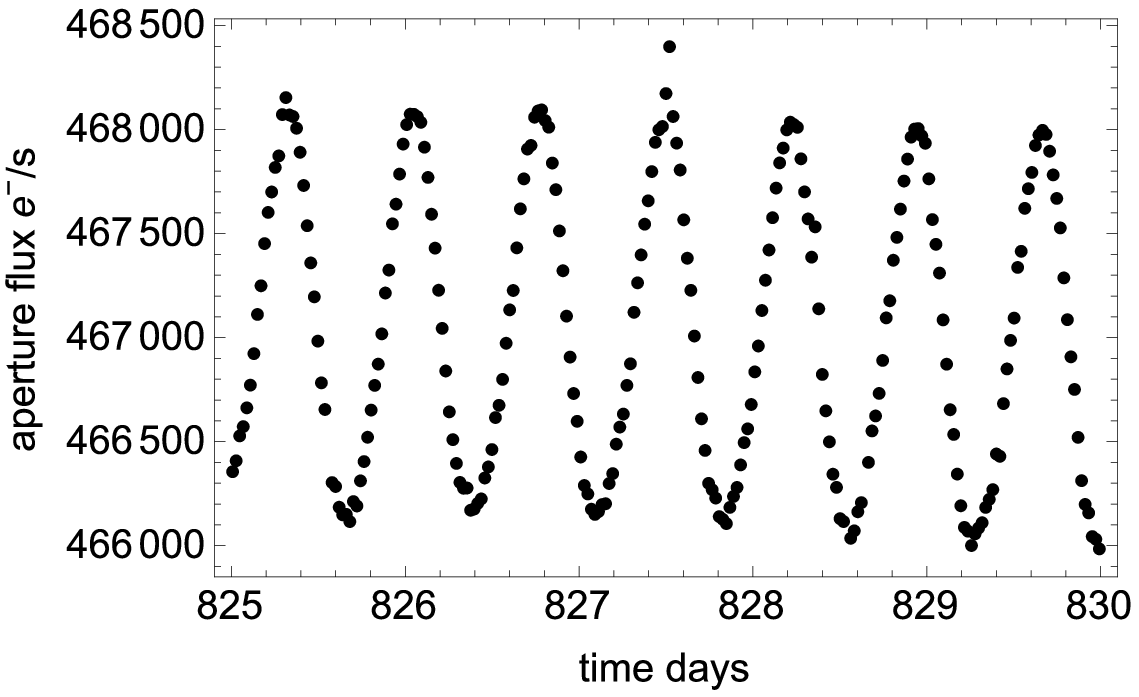}{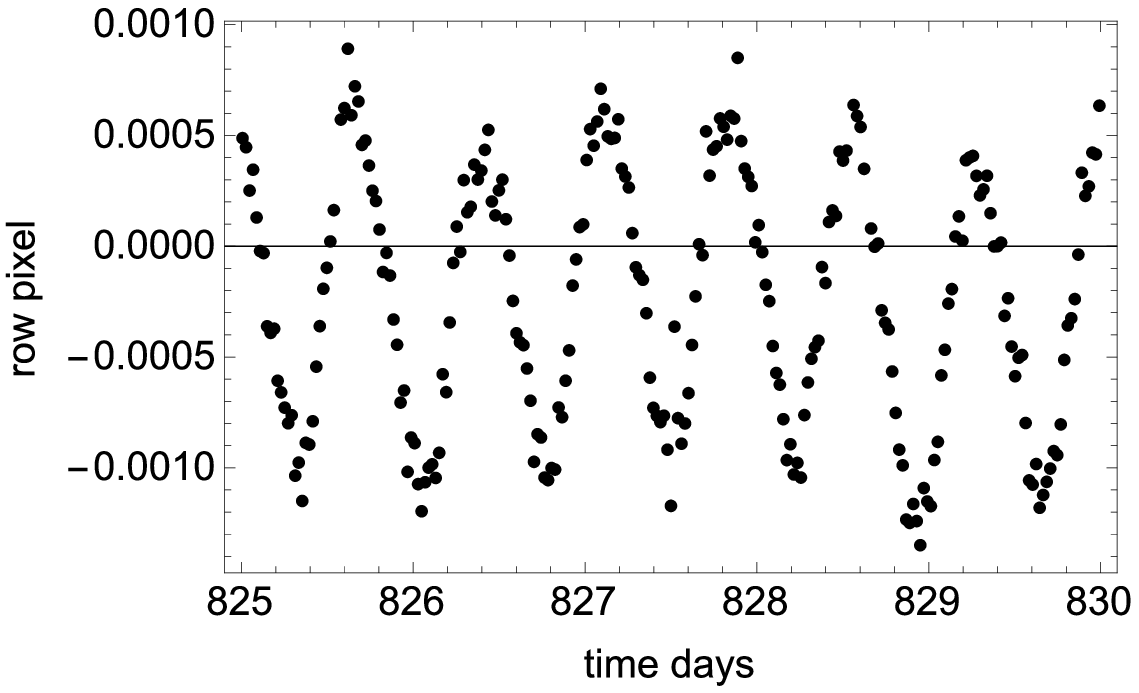}
\hspace{2pc}
\caption{Segments of the long-cadence light curve (left) and row pixel coordinates for the star KIC 3626094
from Quarter 9 data. Despite the marginally significant VIM detection of $r=0.31$ in our catalog, very clear
correlation is seen between the brightness variations and the centroid position.}
\label{362.fig}
\end{figure}
 
This star appears four times in our catalog, for Quarters 8 ($r=0.31$, $posA=217\degr$), Q9 (0.31, $235\degr$),
Q13 (0.33, $238\degr$), and Q17 (0.77, $232\degr$). 
A visual inspection reveals that the object is definitely VIM. Fig. \ref{362.fig} shows a small part of
the light curve between mission day 825 and 830 from Q9 (left panel) and the corresponding segment of the
astrometric position in pixel rows. The astrometric creep was removed from the trajectory for this graph
by fitting a third-degree
polynomial over the entire Quarter. Small as it is in amplitude ($\sim 4$ mas on the sky), the astrometric variation
accurately mirrors that of the recorded flux. From the relative magnitudes of variation in position and flux, using
Eq. \ref{sep.eq}, we estimate the minimum separation between the blended images to be $1\farcs 77$. The low level of
noise in the data and the low degree of chromospheric activity confounds the interpretation by \citet{wic} as
a young, Pleiades-age, star. It is probable that the 0.724 d periodicity comes from the fainter blended companion.
Is it possible that the spikes interpreted as superflares come from the blended image too?

The VIM position angles are fairly steady at $\sim 235\degr$ suggesting that the companion is southwest of the target.
The Sloan Digital Sky Survey maps available through the Simbad CDS web portal show a faint companion separated 
by $\sim 15$ arcsec in this direction. More accurate
estimates are obtained from the data in the 2MASS catalog \citep{cut}: separation $\rho=14\farcs 88$, position 
angle $\theta= 220\degr$. This companion is fainter by more than 6 magnitudes in the NIR $J$ band. At first glance,
it appears impossible that such a widely separated star, which is at least a 100 times fainter than the target star,
can generate spikes of $\sim 0.3$\% of the total flux. However, we have noticed surprisingly powerful VIM effects
from widely separated sources on other occasions. If the distant faint companion is the culprit, a direct photometric
monitoring from the ground should reveal huge spikes in its light curve, caused by a very high level of magnetic
activity. The alternative is that another, unresolved companion is hidden much closer to the target star, and the
matching position angle is a mere coincidence.

\section{Conclusions and future plans}
The empirical and straightforward approach adopted for this work resulted in $343\,298$ high-confidence
VIM detections for $129\,525$ Kepler targets. Almost half of the stars have only one detection out of
17, indicating that many detections are caused by
long-range instrumental contamination effects, acting on the scale of a detector CCD or the entire field of view.
For example, we estimated that $\sim 28$\% of VIM detections are caused by the column illumination anomaly,
which affects observations across an entire CCD. Elevated rates of VIM were found for known eclipsing stars, solar-type
stars with suspected superflares, and known resolved doubles.
We adopted an accommodating threshold for VIM detection ($r>0.3$),
which assures a formal statistical confidence level of 100\%. Thus, only the detections with the largest correlations 
between the flux and the centroid position were included in the catalog. The expected rate of false positives
is zero. The motivation for this overly conservative criterion was to attempt reducing the rate of inverse
VIM effects (MIV) in the output, when the apparent variation in flux is caused by the actual motion of the image
with respect to the pixel aperture.  Still, the resulting list may include a large fraction of MIV, judging from
the low mode of the distribution of number of detections per star, which is equal to 1. Indeed, a double star
with a significantly variable component is expected to be detected as VIM in every quarter irrespective of the
telescope orientation, but only $\sim3$\% of our objects were detected in more than 10 quarters out of 17. The intrinsically less variable sources can still be occasionally detected by our algorithm due to imstrumental
perturbations or a considerable amount of astrometric drift relative to the pixel aperture. 
With these known limitations, the main intended purpose of this work is to facilitate a safe and
reliable identification of candidate exoplanet transit events and other astrophysically significant variability
phenomena, irrespective of the nature of detected perturbations. 

This strategy limits the applicability of our results to the detection of new double stars. The theoretical sensitivity
of the collected data to angular separations as small as 80 mas, matching the most advanced and expensive
imaging and interferometric ground-based techniques, is hard to realize with the adopted method. We are looking into
the possibility of separating long-range instrumental VIM from local, short-range events, by using more advanced
data analysis algorithms, similar in spirit
to the principal component analysis. Instrumental contamination of signal, as opposed
to the blending of unresolved double sources, is likely to affect more than one target at a time. By finding similar
position-light correlations for well separated targets, we hope to identify the artifacts and come up with a much cleaner
sample of candidate double stars.

\label{lastpage}

\end{document}